\documentclass[prb,10pt,twocolumn]{revtex4}

\usepackage{epsfig}
\graphicspath{{/}}

\begin{document}

\title{Fast Exciton Annihilation by Capture of Electrons or Holes by Defects via Auger Scattering in Monolayer Metal Dichalcogenides}

\author{Haining Wang, Jared H. Strait, Changjian Zhang, Weimin Chan, Christina Manolatou, Sandip Tiwari, Farhan Rana}
\affiliation{School of Electrical and Computer Engineering, Cornell University, Ithaca, NY 14853}
\email{fr37@cornell.edu}

\begin{abstract}
The strong Coulomb interactions and the small exciton radii in two-dimensional metal dichalcogenides can result in very fast capture of electrons and holes of excitons by mid-gap defects from Auger processes. In the Auger processes considered here, an exciton is annihilated at a defect site with the capture of the electron (or the hole) by the defect and the hole (or the electron) is scattered to a high energy. In the case of excitons, the probability of finding an electron and a hole near each other is enhanced many folds compared to the case of free uncorrelated electrons and holes. Consequently, the rate of carrier capture by defects from Auger scattering for excitons in metal dichalcogenides can be 100-1000 times larger than for uncorrelated electrons and holes for carrier densities in the $10^{11}$-$10^{12}$ cm$^{-2}$ range. We calculate the capture times of electrons and holes by defects and show that the capture times can be in the sub-picosecond to a few picoseconds range. The capture rates exhibit linear as well as quadratic dependence on the exciton density. These fast time scales agree well with the recent experimental observations~\cite{Shi13,Lagarde14,Korn11,Wangb14}, and point to the importance of controlling defects in metal dichalcogenides for optoelectronic applications.     
\end{abstract}
                                    
\maketitle

\section{Introduction}
Many body interactions play an important role in determining the electronic and optoelectronic properties of two-dimensional (2D) transition metal dichalcogenides (TMDs). The exciton  binding energies in 2D chalcogenides are almost an order of magnitude larger compared to other bulk semiconductors \cite{Fai10,Xu13,Changjian14,timothy,Chernikov14}. The strong Coulomb interactions and small exciton radii in 2D-TMDs result in large optical oscillator strengths~\cite{Changjian14,Konabe14,Berg14} and short radiative lifetimes~\cite{Wanga14}. In this paper we show that the same factors also result in very fast capture of electrons and holes of excitons by defects from Auger processes leading to fast non-radiative recombination rates. The basic idea can be understood as follows. Consider the Auger process in which a hole (in the valence band) scatters off an electron (in the conduction band) and is captured by a mid-gap defect level and the electron (in the conduction band) takes the energy released in the hole capture process. In the case of uncorrelated electrons and holes, the rate for this process is proportional to the product of the hole density $p$ and the probability of finding an electron near the hole, which is proportional to the electron density $n$. But in the case of tightly bound excitons, an electron is present near the hole with a very high probability proportional to $|\phi(\vec{r}=0)|^{2}$, where $\phi(\vec{r})$ is related to the exciton wavefunction (see the discussion below). Therefore, the rate for a hole (or an electron) in a tightly bound exciton to get captured by a defect is proportional to the exciton density times $|\phi(\vec{r}=0)|^{2}$. Generally speaking, Auger rates in semiconductors are considered to be important only at large carrier densities~\cite{Landsberg92}. But given the small exciton radii in 2D-TMDs (in the 7-10$\AA$ range), $|\phi(\vec{r}=0)|^{2}$, which is inversely proportional to the square of the exciton radius, can be extremely large and, consequently, Auger capture rates in 2D-TMDs can be very fast. Compared to the rates for direct electron-hole recombination via interband Auger scattering (exciton-exciton annihilation), which can be limited by the orthogonality of the conduction and valence band Bloch states, the rates for the capture of electrons and holes of excitons by defects can be very fast when the defect states have a good overlap with the conduction or valence band Bloch states.

Quantum efficiencies of TMD light emitters and detectors that have been reported are extremely poor; in the .0001-.01 range~\cite{Lopez13, Ross14,Yin12,Steiner13,Pablo14}. Similar quantum efficiencies for TMDs have been observed in photoluminescence experiments~\cite{Fai10,Feng10,Wangb14}. Therefore, most of the electrons and holes injected electrically or optically in TMDs recombine non-radiatively. Given that the average radiative lifetimes of excitons in TMDs are in the range of hundreds of picoseconds to a few nanoseconds~\cite{Wanga14}, the non-radiative recombination or capture times in TMDs are expected to be of the order of a few picoseconds. Several experimental results on the ultrafast carrier dynamics in photoexcited monolayer MoS$_{2}$ do indeed point to non-radiative recombinaton and/or capture times in the few picoseconds range~\cite{Shi13,Korn11,Lagarde14,Wangb14}. The mechanisms by which electrons and holes recombine non-radiatively and/or are captured by defects, and the associated time scales, remain to be clarified. The results in this paper show that electrons and holes of excitons in TMDs can get captured by defects on very short times scales that are in the sub-picosecond to a few picoseconds range resulting in exciton annihilation. The capture rates exhibit linear as well as quadratic dependence on the exciton density. The quadratic dependence of the exciton annihilation rate on the exciton density is generally considered to be an exclusive characteristic of exciton-exciton annihilation processes via interband Auger scattering. Although the discussion in this paper focuses on monolayer MoS$_{2}$, the analysis and the results presented here are expected to be relevant to all 2D-TMDs, and are expected to be useful in designing metal dichalcogenide optoelectronic devices as well as in helping to understand and interpret experimental data~\cite{Shi13,Lagarde14,Korn11,Wangb14}.

\section{Theoretical Model} 
\begin{figure}[tbh]
  \centering
  \includegraphics[width=.3\textwidth]{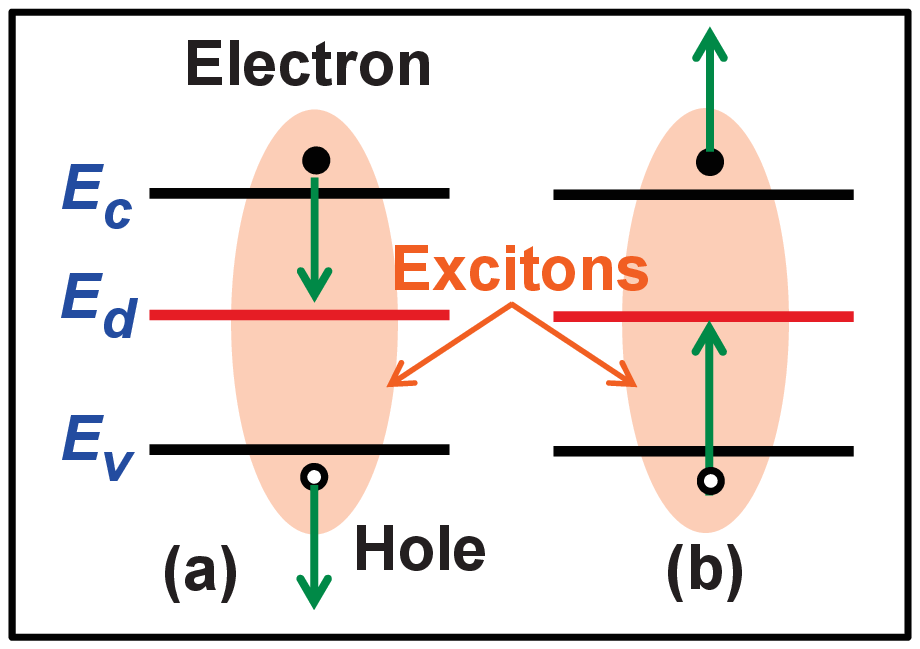}
  \caption{Two basic Auger processes for the capture of an electron (a) or a hole (b) of an exciton by a defect state are depicted~\cite{Landsberg92,Landsberg80}.}
  \label{fig:auger_1}
\end{figure}
\subsection{Introduction}
The two basic Auger processes for the capture of an electron (a) or a hole (b) of an exciton by a defect state are depicted in Fig.\ref{fig:auger_1}. Proper partitioning of the Hamiltonian is important in order to compute the rates of these processes. We discuss the terms in the Hamiltonian describing various processes below. 

\subsection{The Non-interacting Hamiltonian} \label{sec:nonintH}
The crystal structure of a monolayer of group-VI dichalcogenides $MX_{2}$ (e.g. $M$=Mo,W and $X$=S,Se) consist of $X$-$M$-$X$ layers, and within each layer the $M$ atoms (or the $X$ atoms) form a 2D hexagonal lattice. Each $M$ atoms is surrounded by 6 nearest neighbor $X$ atoms in a trigonal prismatic geometry with $D_{3h}^{1}$ symmetry. The valence band maxima and conduction band minima occur at the $K$ and $K'$ points in the Brillouin zone. Most of the weight in the conduction and valence band Bloch states near the $K$ and $K'$ points resides on the d-orbitals of $M$ atoms~\cite{Falko13,yao12,timothy}. The spin up and down valence bands are split near the  $K$ and $K'$ points by 0.1-0.2 eV due to the spin-orbit-coupling~\cite{Falko13,yao12,timothy,Liu14}. In comparison, the spin-orbit-coupling effects in the conduction band are much smaller~\cite{Liu14}. Assuming only d-orbitals for the conduction and valence band states, and including spin-orbit coupling, one obtains the following simple spin-dependent tight-binding Hamiltonian (in matrix form) near the $K$($K'$) points \cite{yao12},
\begin{equation}
\left[
\begin{array}{cc}
\Delta/2 & \hbar v k_{-} \\
\hbar v k_{+} & -\Delta/2 + \lambda \tau \sigma
\end{array} \right]    \label{eq:H1}
\end{equation}
Here, $\Delta$ is related to the material bandgap, $\sigma=\pm1$ stands for the electron spin, $\tau=\pm1$ stands for the $K$ and $K'$ valleys, $2\lambda$ is the splitting of the valence band due to spin-orbit coupling, $k_{\pm}=\tau k_{x}\pm ik_{y}$, and the velocity parameter $v$ is related to the coupling between the orbitals on neighboring $M$ atoms. From density functional theories \cite{Lam12,Falko13}, $v\approx 5-6 \times 10^5 $ m/s. The wavevectors are measured from the $K$($K'$) points. The d-orbital basis used in writing the above Hamiltonian are $| d_{z^{2}} \rangle$ and $(| d_{x^{2}-y^{2}}\rangle + i\tau | d_{xy}\rangle)/\sqrt{2}$~\cite{yao12}. We will use the symbol $s$ for the combined valley ($\tau$) and spin ($\sigma$) degrees of freedom. Defining $\Delta_{s}$ as $\Delta - \lambda \tau \sigma$, the energies and eigenvectors of the conduction and valence bands are~\cite{yao12,Efimkin13},
\begin{equation}
E_{{c \atop v},s}(\vec{k}) = \frac{\lambda \tau \sigma}{2} + \gamma \sqrt{ (\Delta_{s}/2)^{2} + (\hbar v k)^{2}}
\end{equation}
\begin{equation}
| v_{{c \atop v},\vec{k},s}\rangle  = \left[ \begin{array}{c} \cos(\theta_{\gamma,\vec{k},s}/2) e^{-i\tau\phi_{\vec{k}}/2} \\ \tau \gamma \sin(\theta_{\gamma,\vec{k},s}/2) e^{i\tau\phi_{\vec{k}}/2} \end{array} \right] \label{eq:wf1}
\end{equation}
Here, $\gamma =1$ (or $-1$) stands for the conduction (or the valence) band, $\phi_{\vec{k}}$ is the phase of the wavevector $\vec{k}$, and,
\begin{equation}
\cos(\theta_{\gamma,\vec{k},s}) = \gamma \frac{\Delta_{s}}{2 \sqrt{ (\Delta_{s}/2)^{2} + (\hbar v k)^{2}}} \label{eq:cs}
\end{equation}
Near the conduction band minima and valence band maxima, the band energy dispersion is parabolic with well-defined effective masses, $m_{e}$ and $m_{h}$, for electrons and holes, respectively. 

The Hamiltonian describing electron states in the conduction band, valence band, and a mid-gap defect state is,
\begin{eqnarray}
H_{o} & = & \sum_{\vec{k},s}E_{c,s}(\vec{k}) c^{\dagger}_{\vec{k},s} c_{\vec{k},s} + \sum_{\vec{k},s}E_{v,s}(\vec{k}) b^{\dagger}_{\vec{k},s} b_{\vec{k},s} \nonumber \\
& & + \sum_{\sigma}E_{d} d^{\dagger}_{\sigma}d_{\sigma}
\end{eqnarray}
Here, $c_{\vec{k},s}$, $b_{\vec{k},s}$, and  $d_{\sigma}$ are the destruction operators for the conduction band, valence band, and defect states, respectively. The bandgap is $E_{g_{s,s'}} = E_{c,s}(\vec{k}=0) - E_{v,s'}(\vec{k}=0)$. Since only the smallest bandgap will be relevant in the discussion that follows, we will drop the spin/valley indices from $E_{g_{s,s'}}$ for simplicity.

\subsection{Electron-Hole Interaction and Exciton States} \label{sec:Heh}
The Coulomb interaction between the electrons and holes can be included by adding the following term to the Hamiltonian,
\begin{equation}
H_{eh} =  \frac{1}{A} \sum_{\vec{k},\vec{k}',\vec{q},s,s'} V(\vec{q})  F_{s,s'}(\vec{k},\vec{k}',\vec{q}) c^{\dagger}_{\vec{k}+\vec{q},s}  b^{\dagger}_{\vec{k}'-\vec{q},s'} b_{\vec{k}',s'} c_{\vec{k},s}
\end{equation}
$V(\vec{q})$ is the 2D Coulomb potential and equals $e^{2}/2\epsilon_{o} \epsilon(\vec{q})q$. The wavevector-dependent dielectric constant $\epsilon(\vec{q})$ for monolayer MoS$_{2}$ is given by Zhang et~al.~\cite{Changjian14} and  Berkelbach et~al.~\cite{timothy}. $F_{s,s'}(\vec{k},\vec{k}',\vec{q})$ is~\cite{Efimkin13},
\begin{equation}
F_{s,s'}(\vec{k},\vec{k}',\vec{q}) = \langle v_{c,\vec{k}+\vec{q},s}| v_{c,\vec{k},s} \rangle \, \langle v_{v,\vec{k}'-\vec{q},s'} | v_{v,\vec{k}',s'} \rangle \label{eq:F1}
\end{equation}
Near the conduction band minima, where $\hbar v k << \Delta_{s}$, $\cos(\theta_{\gamma,\vec{k},s}) \approx 1$ and $\sin(\theta_{\gamma,\vec{k},s}) << 1$. Similarly, near the valence band maxima, $\sin(\theta_{\gamma,\vec{k},s}) \approx 1$ and  $\cos(\theta_{\gamma,\vec{k},s}) << 1$. Therefore, for wavevectors near the band extrema one can make the approximation~\cite{Efimkin13},
\begin{equation}
F_{s,s'}(\vec{k},\vec{k}',\vec{q}) = e^{i(\tau\phi_{\vec{k}+\vec{q}} - \tau\phi_{\vec{k}} + \tau'\phi_{\vec{k}'} - \tau'\phi_{\vec{k}'-\vec{q}})/2} \label{eq:F2}
\end{equation}
Exciton states are approximate eigenstates of the Hamiltonian $H_{o} + H_{eh}$. Assuming that the ground state of the semiconductor is $|\psi_{o}\rangle$, which consists of a filled valence band and an empty conduction band, an exciton state with in-plane momentum $\vec{Q}$ can be constructed from the ground state as follows~\cite{Changjian14,Efimkin13}, 
\begin{equation}
|\psi_{s,s',\alpha}(\vec{Q})\rangle =   \frac{1}{\sqrt{A}}\sum_{\vec{k}} \psi_{\alpha,\vec{Q}}(\vec{k}) c_{\vec{k} + \frac{m_{e}}{m_{ex}}\vec{Q},s}^{\dagger} b_{\vec{k}-\frac{m_{h}}{m_{ex}}\vec{Q},s'}|\psi_{o} \rangle  \label{eq:exciton}
\end{equation} 
The exciton wavefunction is $\psi_{\alpha,\vec{Q}}(\vec{k})$. The electron and hole effective masses are $m_{e}$ and $m_{h}$, respectively. The exciton mass is $m_{ex} = m_{e} + m_{h}$, and the reduced electron-hole mass is $m_{r}$. If one writes the exciton wavefunction as,
\begin{equation}
\psi_{\alpha,\vec{Q}}(\vec{k}) = \phi_{\alpha}(\vec{k}) e^{i(\tau\phi_{\vec{k}+ (m_{e}/m_{ex})\vec{Q}} + \tau'\phi_{\vec{k}'-(m_{h}/m_{ex})\vec{Q}})/2} \label{eq:psiphi}
\end{equation}
then the exciton wavefunction $\phi_{\alpha}(\vec{k})$ satisfies the standard exciton eigenvalue equation~\cite{Changjian14,timothy},
\begin{eqnarray}
& & \left[\bar{E}_{c}(\vec{k}) - \bar{E}_{v}(\vec{k}) \right]\phi_{\alpha}(\vec{k}) - \frac{1}{A} \sum_{\vec{q}} V(\vec{q}) \phi_{\alpha}(\vec{k}-\vec{q}) \nonumber \\
& & = E_{\alpha}(\vec{Q}) \phi_{\alpha}(\vec{k}) \label{eq:eigen}
\end{eqnarray}
with an eigenvalue $E_{\alpha}(\vec{Q})$ given by,
\begin{equation}
E_{\alpha}(\vec{Q}) = E_{g} - E_{\alpha} + \frac{\hbar^{2} Q^{2}}{2 m_{ex}}
\end{equation}
where, $E_{\alpha}$ is the exciton binding energy. The energy $E_{\alpha}(\vec{Q})$ is measured with respect to the energy of the ground state $|\psi_{o}\rangle$. {\em Note that the phase factors cancel out and do not appear in the exciton eigenvalue equation}. The exciton wavefunctions are orthonormal and complete in the sense~\cite{Kira12}, 
\begin{equation}
\int \frac{d^{2}\vec{k}}{(2\pi)^{2}} \phi^{*}_{\alpha}(\vec{k}) \phi_{\beta}(\vec{k}) = \delta_{\alpha,\beta} \label{eq:orth} 
\end{equation}
\begin{equation}
\sum_{\alpha} \phi_{\alpha}(\vec{k}) \phi^{*}_{\alpha}(\vec{k}') = (2\pi)^{2} \delta^{2}(\vec{k}-\vec{k}') \label{eq:comp}
\end{equation}
The sum over $\alpha$ above includes all the discrete bound exciton states as well as the continuum of ionized exciton states. Finally, the probability of finding an electron and a hole at a distance $\vec{r}$ in the exciton state $| \psi_{\alpha,\vec{Q}}(\vec{k}) \rangle$ can be computed by destroying an electron and a hole using the real-space field destruction operators and then taking the overlap of the resulting state with the ground state $|\psi_{o}\rangle$. The result is $|\phi_{\alpha}(\vec{r})|^{2}$ where $\phi_{\alpha}(\vec{r})$ is the Fourier transform of $\phi_{\alpha}(\vec{k})$. Note that $\phi_{\alpha}(\vec{r})$ is not the Fourier transform of $\psi_{\alpha,\vec{Q}}(\vec{k})$, which also includes extra phase factors (see (\ref{eq:psiphi}).

\subsection{Exciton Basis} \label{sec:ex_basis}
In what follows, we will use the exciton basis. The exciton creation operator $B^{\dagger}_{s,s',\alpha}(\vec{Q})$ can be defined as,
\begin{equation}
 B^{\dagger}_{s,s',\alpha}(\vec{Q}) =  \frac{1}{\sqrt{A}}\sum_{\vec{k}} \psi_{\alpha,\vec{Q}}(\vec{k}) c_{\vec{k} + \frac{m_{e}}{m_{ex}}\vec{Q},s}^{\dagger} b_{\vec{k}-\frac{m_{h}}{m_{ex}}\vec{Q},s'} 
\end{equation}
Using the completeness and the orthogonality of the exciton wavefunctions given in (\ref{eq:comp}) and (\ref{eq:orth}), we get,
\begin{equation}
c_{\vec{k},s}^{\dagger} b_{\vec{k}',s'} = \frac{1}{\sqrt{A}}\sum_{\alpha} \psi^{*}_{\alpha,\vec{Q}}(\vec{k}_{r}) B^{\dagger}_{s,s',\alpha}(\vec{Q})
\end{equation}
Here, $\vec{k}_{r}$ and $\vec{Q}$ equal $(m_{h}/m_{ex})\vec{k} + (m_{e}/m_{ex})\vec{k}'$ and $\vec{k}-\vec{k}'$ on the left hand side, respectively. Products of electron and hole creation and destruction operators can thus be expressed in terms of the exciton operators.  

\begin{figure}[tbh]
  \centering
  \includegraphics[width=.45\textwidth]{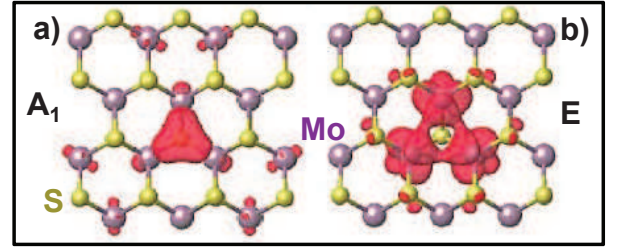}
  \caption{The computed orbitals of the defect states in a MoS$_{2}$ monolayer corresponding to a sulfur vacancy are shown (from Noh et~al.~\cite{Noh14}). (a) and (b) show $A_{1}$ state and the two degenerate $E$ states, respectively.}
  \label{fig:s_vac}
\end{figure}

\subsection{Defect States}
TMDs ($MX_{2}$), and in particular Monolayer MoS$_{2}$, are known to have several different kinds of point defects, such as $M$ and $X$ vacancies and interstitials, impurity atoms, in addition to grain boundaries and dislocations~\cite{Sofo04,Komsa12,Seifert13,Kong13,Noh14,Guinea14,Robertson13,Hao13,VanDerZande13}. The goal in this Section is not to give a detailed description of different defect states in TMDs, something well beyond the scope of this paper, but to capture the essential physics in a way that would enable us to obtain capture rates for electrons and holes and present the main ideas associated with the capture processes. 

Since the Bloch states form a complete set, the wavefunction $\psi_{d}(\vec{r})$ of the electron in the defect state can be expanded in terms of the Bloch states from all the bands~\cite{Landsberg92}. In most cases of practical interest, only Bloch states in the vicinity of certain points, $\vec{K}_{s}$, in the Brillouin zone, such as $\Gamma$, $M$, $K$ and $K'$ in the case of 2D-TMDs, need to be included in the expansion and therefore one may write,
\begin{equation}
\psi_{d}(\vec{r}) = \frac{1}{\sqrt{A}} \sum_{n,\vec{k},s} c_{n,s}(\vec{k}) \frac{e^{i(\vec{K}_{s} + \vec{k}).\vec{r}}}{\sqrt{A}} u_{n,\vec{k},s}(\vec{r}) \label{eq:defect1}
\end{equation}
In the expression above, $u_{n,\vec{k},s}(\vec{r})$ are the periodic parts of the Bloch functions. The sum over $n$ runs over all the energy bands. Whereas shallow defect levels can usually be described well by limiting the summation above to a single band, deep mid-gap defect levels generally have contributions from multiple bands~\cite{Landsberg92,Landsberg80}. The above expression can usually be cast in much simpler forms for specific defect states. 

As an example, we consider the case of the deep point defect in MoS$_{2}$ due to a sulfur atom vacancy. A sulfur atom vacancy is a common defect in MoS$_{2}$ monolayers and can have a small formation energy~\cite{Kong13,Robertson13,Noh14}. The three states within the bandgap associated with a sulfur vacancy have been obtained previously using ab-initio techniques~\cite{Kong13,Robertson13,Noh14}. These defect states consist of: (i) a single $A_{1}$ state, made up of mostly the $d_{xz}$ and $d_{yz}$ orbitals of the Mo atoms adjacent to the missing S atom, with an energy few tenths of an eV above the valence band maxima, and (ii) two degenerate $E$ states, made up of mostly the $d_{z^{2}}$, $d_{x^{2}-y^{2}}$, and $d_{xy}$ orbitals of the Mo atoms adjacent to the missing S atom, with an energy 1.4-1.6 eV above the valence band maxima. All the defect states are spin-degenerate and correspond to the one ($A_{1}$) and two dimensional ($E$) representations of the trigonal symmetry group $C_{3v}$. The computed orbitals of these states are shown in Fig.\ref{fig:s_vac} (from Noh et~al.~\cite{Noh14}). A defect state can be an efficient center for non-radiative recombination due to Auger scattering only if it has good overlaps with the Bloch states of both the conduction and the valence bands. The $E$ states fit this criterion. The $E$ states can be described well by limiting the summation in the expression above to the Bloch states of the conduction and the valence band extrema at the $K$ and $K'$ points. Since all the orbitals forming the $E$ states have weights almost entirely on the Mo atoms adjacent to the missing S atom, one may write $c_{{c \atop v},s}(\vec{k}) \approx \chi_{d}(\vec{k}) e^{i\gamma\tau\phi_{\vec{k}}/2} b_{{c \atop v},s}$. Since $e^{i\gamma\tau\phi_{\vec{k}}/2}  u_{{c \atop v},\vec{k},s}(\vec{r})$ does not vary much with $\vec{k}$ near the band extrema, the sum in (\ref{eq:defect1}) can be rearranged to give,
\begin{equation}
\psi_{d}(\vec{r}) =  \chi_{d}(\vec{r}) \sum_{{n=c,v \atop s}} b_{n,s} e^{i\vec{K}_{s}.\vec{r}} e^{i\gamma\tau\phi_{\underline{\vec{k}}}/2} u_{n,\underline{\vec{k}},s}(\vec{r}) \label{eq:defect2}
\end{equation}
Here, the line under $\vec{k}$ means that any wavevector near the band extrema can be chosen. The function $\chi_{d}(\vec{r})$ is expected to be localized at the defect, becoming very small at the second nearest Mo atom near the defect site.

\subsection{Hamiltonian for the Capture of Holes and Electrons} \label{sec:Hc}
Consider process (b) in Fig.\ref{fig:auger_1} in which a hole scatters off an electron and is captured by a defect and the electron is scattered to a higher energy. The relevant term in the Coulomb interaction Hamiltonian that describes the hole capture process in Fig.\ref{fig:auger_1}(b) can be written as,
\begin{equation}
H_{hc} =  \frac{1}{A} \sum_{\vec{k},\vec{k}',\vec{q},s,s'} V(\vec{q}) M_{s,s'}(\vec{k},\vec{k}',\vec{q}) c^{\dagger}_{\vec{k}+\vec{q},s} b^{\dagger}_{\vec{k}',s'} d_{\sigma'} c_{\vec{k},s}   + h.c.
\end{equation}
The overlap factor $M_{s,s'}(\vec{k},\vec{k}',\vec{q})$ equals,
\begin{eqnarray}
& & M_{s,s'}(\vec{k},\vec{k}',\vec{q})  =  \langle v_{c,\vec{k}+\vec{q},s} | v_{c,\vec{k},s} \rangle \nonumber \\
& & \times  \sum_{n=c,v} b_{n,s'} e^{i\gamma\tau'\phi_{\underline{\vec{k}}}/2} \langle v_{v,\vec{k}',s'} | v_{n,\underline{\vec{k}},s'} \rangle \nonumber \\
& & \times \frac{1}{\sqrt{A}} \int d^{2}\vec{r} \, \chi_{d}(\vec{r}) \, e^{-i(\vec{k}' + \vec{q}).\vec{r}} \nonumber \\
& & \approx    e^{i(\tau\phi_{\vec{k}+\vec{q}} - \tau\phi_{\vec{k}}  - \tau'\phi_{\vec{k}'})/2} \frac{b_{v,s'}}{\sqrt{A}} \chi_{d}(\vec{k}'+\vec{q})
\end{eqnarray}
Similarly, the electron capture process (Fig.\ref{fig:auger_1}(a)) is described by the Hamiltonian,
\begin{equation}
H_{ec} =  \frac{1}{A} \sum_{\vec{k},\vec{k}',\vec{q},s,s'} V(\vec{q}) L_{s,s'}(\vec{k},\vec{k}',\vec{q})  d^{\dagger}_{\sigma} b^{\dagger}_{\vec{k}',s'} b_{\vec{k}'+\vec{q},s'} c_{\vec{k},s}  +  h.c.
\end{equation}
where overlap factor $L_{s,s'}(\vec{k},\vec{k}',\vec{q})$ equals,
\begin{equation}
L_{s,s'}(\vec{k},\vec{k}',\vec{q}) \approx   e^{i(\tau'\phi_{\vec{k}'+\vec{q}} - \tau'\phi_{\vec{k}'}  - \tau \phi_{\vec{k}})/2} \frac{b^{*}_{c,s}}{\sqrt{A}} \chi^{*}_{d}(\vec{k}+\vec{q}) 
\end{equation}
The potential of the defect does not appear in the Hamiltonian above. The reason for this is that it has already been taken into account in defining the non-interacting Hamiltonian, and its eigenstates, in Section (\ref{sec:nonintH}).

\section{Electron and Hole Capture Rates for Excitons}
We assume an initial state described by the density operator $\rho_{i}$ in which the exciton occupation $n_{s,s',\alpha}(\vec{Q})$, defect occupation $f_{d}$, and conduction and valence band occupations are given by,
\begin{eqnarray}
&& \langle d^{\dagger}_{\sigma}  d_{\sigma'}  \rangle    =  f_{d} \, \delta_{\sigma,\sigma'}  \nonumber \\ \label{eq:ex_avg}
&& \langle  c^{\dagger}_{\vec{k},s}  c_{\vec{k}',s'}\rangle  = f_{c,s}(\vec{k})  \delta_{s,s'}  \delta_{\vec{k},\vec{k}'} \nonumber \\
&& \langle   b^{\dagger}_{\vec{k},s}  b_{\vec{k}',s'}  \rangle  = f_{v,s}(\vec{k}) \delta_{s,s'}  \delta_{\vec{k},\vec{k}'} \nonumber \\
&& \langle B^{\dagger}_{s,s',\alpha}(\vec{Q})  B_{s,s',\alpha}(\vec{Q}) \rangle   =  n_{s,s',\alpha}(\vec{Q})  + \nonumber \\
&& \frac{1}{A}\sum_{\vec{k}} |\phi_{\alpha}(\vec{k})|^{2} f_{c,s}(\vec{k} + \frac{m_{e}}{m_{ex}} \vec{Q}) \left[ 1 - f_{v,s'}(\vec{k} -  \frac{m_{h}}{m_{ex}} \vec{Q}) \right] \nonumber \\ 
\end{eqnarray}
The angled brackets stand for ensemble averaging with respect to the the density operator $\rho_{i}$. Since the excitons are not exact bosons, the value of $\langle B^{\dagger}_{s,s',\alpha}(\vec{Q})  B_{s,s',\alpha}(\vec{Q}) \rangle$ is not just equal to the exciton occupation $n_{s,s',\alpha}(\vec{Q})$. Using the cluster expansion to evaluate $\langle B^{\dagger}_{s,s',\alpha}(\vec{Q})  B_{s,s',\alpha}(\vec{Q}) \rangle$ results in the additional Hartree-Fock term shown above~\cite{Kira06,Koch06}. The same extra term also shows up in the luminescence spectra of excitons~\cite{Kira12}, and, as discussed below, this term results in a quadratic dependence of the capture rate on the exciton density at large exciton densities. 

We assume that the electron and hole densities for different spins/valleys (including both free carriers and bound excitons) are $n_{s}$ and $p_{s'}$, respectively, and the defect density is $n_{d}$. The initial ensemble consists of states that are approximate eigenstates of $H_{o} + H_{eh}$ but not of $H_{o} + H_{eh} + H_{hc} + H_{ec}$. Therefore, we consider $H_{hc}$ and $H_{ec}$ as perturbations.

\subsection{Electron Capture Rate} 
We first consider process (a) in Fig.\ref{fig:auger_1} in which the electron is captured by a defect. The average electron capture rate $R_{ec}$ (units: per unit area per second) can be calculated from the first order perturbation theory using the exciton basis described in Section~\ref{sec:ex_basis} and the average values given in (\ref{eq:ex_avg}). The details of the calculations are given in the Appendix. The final result is,
\begin{eqnarray}
& & R_{ec} \approx \frac{2\pi}{\hbar} n_{d}(1-f_{d}) \sum_{s,s',\alpha} D_{v,s'}(q_{\alpha}) |\chi_{d}(q_{\alpha})|^{2}  |b_{c,s}|^{2}   \nonumber \\ 
& & \times \left| \frac{1}{A} \sum_{\vec{k}_{r}} V(q_{\alpha}\hat{x}-\vec{k}_{r})  \phi_{\alpha}(\vec{k}_{r}) \right|^{2} \left[ n_{s,s',\alpha} \right. \nonumber \\
& & \left. + \frac{1}{A^{2}} \sum_{\vec{k},\vec{Q}} |\phi_{\alpha}(\vec{k})|^{2} f_{c,s}(\vec{k} + \frac{m_{e}}{m_{ex}} \vec{Q}) \left[ 1 - f_{v,s'}(\vec{k} -  \frac{m_{h}}{m_{ex}} \vec{Q}) \right] \right] \nonumber \label{eq:Rec_0} \\ 
\end{eqnarray}
Here, $D_{v,s'}$ is the valence band density of states (per valley per spin) evaluated at the energy of the scattered hole whose wavevector is $q_{\alpha}$. $q_{\alpha}$ is approximately given by the relation, $E_{v,s'}(0) - E_{v,s'}(q_{\alpha}) = E_{g} - E_{\alpha} - E_{d}$. Note that none of the phase factors appear in the above result. The exciton density $n_{s,s',\alpha}$ is, 
\begin{equation}
n_{s,s',\alpha} = \int \frac{d^{2}\vec{Q}}{(2\pi)^{2}} \, n_{s,s',\alpha}(\vec{Q})
\end{equation} 
If $q_{\alpha} >> k_{r}$ for all values of $k_{r}$ for which $\phi_{\alpha}(\vec{k}_{r})$ is significant, then the above expression reduces to, 
\begin{eqnarray}
R_{ec} & = & \frac{2\pi}{\hbar} n_{d} \left( 1-f_{d} \right) \sum_{s,s',\alpha} D_{v,s'}(q_{\alpha}) |V(q_{\alpha})|^{2} |\chi_{d}(q_{\alpha})|^{2}  \nonumber \\
& &  \times |b_{c,s}|^{2} \left[ |\phi_{\alpha}(\vec{r}=0)|^{2} n_{s,s',\alpha} + G_{\alpha} n_{s}p_{s'} \right]         \label{eq:Rec_1}
\end{eqnarray}
Expression for $G_{\alpha}$ is given in the Appendix. $G_{\alpha}$ is significant for only the lowest few exciton states.

\subsection{Hole Capture Rate} 
The rate for process (b) in Fig.\ref{fig:auger_1} in which the hole is captured by a defect can be calculated in the same way. The result is,
\begin{eqnarray}
& & R_{hc} \approx \frac{2\pi}{\hbar} n_{d}f_{d} \sum_{s,s',\alpha} D_{c,s}(q_{\alpha}) |\chi_{d}(q_{\alpha})|^{2}  |b_{v,s'}|^{2}   \nonumber \\
& & \times \left| \frac{1}{A} \sum_{\vec{k}_{r}} V(q_{\alpha}\hat{x}-\vec{k}_{r})  \phi_{\alpha}(\vec{k}_{r}) \right|^{2} \left[ n_{s,s',\alpha} \right. \nonumber \\
& & \left. + \frac{1}{A^{2}} \sum_{\vec{k},\vec{Q}} |\phi_{\alpha}(\vec{k})|^{2} f_{c,s}(\vec{k} + \frac{m_{e}}{m_{ex}} \vec{Q}) \left[ 1 - f_{v,s'}(\vec{k} -  \frac{m_{h}}{m_{ex}} \vec{Q}) \right] \right] \nonumber \\ \label{eq:Rhc_0}
\end{eqnarray}
where now $q_{\alpha}$ is approximately given by the relation, $E_{c,s}(q_{\alpha}) - E_{c,s}(0) = E_{d} - E_{\alpha}$. And, as before, if $q_{\alpha} >> k_{r}$ for all values of $k_{r}$ for which $\phi_{\alpha}(\vec{k}_{r})$ is significant, then the above expression reduces to, 
\begin{eqnarray}
R_{hc} & = & \frac{2 \pi}{\hbar} n_{d} f_{d} \sum_{s,s',\alpha} D_{c,s}(q_{\alpha}) |V(q_{\alpha})|^{2} |\chi_{d}(q_{\alpha})|^{2}  \nonumber \\
& &  \times |b_{v,s'}|^{2} \left[ |\phi_{\alpha}(\vec{r}=0)|^{2} n_{s,s',\alpha} + G_{\alpha} n_{s}p_{s'} \right]       \label{eq:Rhc_1}
\end{eqnarray}

\subsection{Coulomb Correlations and Enhancement of the Auger Capture Rates} \label{sec:corr}
Equation (\ref{eq:Rec_1}) for the electron capture rate can also be written as,
\begin{eqnarray}
R_{ec} & = & \frac{2\pi}{\hbar} n_{d} \left( 1-f_{d} \right) \sum_{s,s',\alpha}  D_{v,s'}(q_{\alpha}) |V(q_{\alpha})|^{2} |\chi_{d}(q_{\alpha})|^{2}  \nonumber \\
& &  \times |b_{c,s}|^{2}  n_{s}p_{s'} \left[  G_{\alpha} + g_{s,s',\alpha}(\vec{r}=0) \right]    \label{eq:Rec_2}
\end{eqnarray}
where, $g_{s,s',\alpha}(\vec{r}=0) = |\phi_{\alpha}(\vec{r}=0)|^{2} n_{s,s',\alpha}/(n_{s}p_{s'})$. The quantity inside the square brackets in (\ref{eq:Rec_2}), $G_{\alpha} + g_{s,s',\alpha}(\vec{r}=0)$, describes the enhancement in the probability of finding an electron and a hole close to each other as a result of the attractive Coulomb interactions. It is interesting to compare the electron capture rate in (\ref{eq:Rec_2}) with the result obtained assuming no electron-hole attractive interaction (i.e. $H_{eh}=0$),
\begin{eqnarray}
R_{ec} & = & \frac{2\pi}{\hbar} n_{d} \left( 1-f_{d} \right) \nonumber \\  
& \times & \sum_{s,s'} D_{v,s'}(q_{o}) |V(q_{o})|^{2} |\chi_{d}(q_{o})|^{2}  |b_{c,s}|^{2} n_{s} p_{s'} \label{eq:Rec_3}
\end{eqnarray}
where $q_{o}$ is approximately given by the relation, $E_{v,s'}(0) - E_{v,s'}(q_{o}) = E_{g} - E_{d}$. It can be seen that the capture rate in (\ref{eq:Rec_2}) is larger by the same enhancement factor. Assuming all the electrons and holes are in the lowest ($\alpha =1$) bound exciton state, values of $D_{v,s'}$ and $|b_{c,s}|$ are independent of the valley/spin indices, and the exciton density is $n_{ex} = \sum_{s,s'} n_{s,s',\alpha=1}$, the comparison between (\ref{eq:Rec_2}) and (\ref{eq:Rec_3}) shows that the enhancement of the electron capture rate in the case of excitons is roughly proportional to $G_{\alpha=1} + |\phi_{\alpha=1}(\vec{r}=0)|^{2}/n_{ex}$. Given that the radius of the lowest exciton state in monolayer MoS$_{2}$ is in the 7-10 $\AA$ range~\cite{Changjian14}, the enhancement, assuming an exciton density of $10^{12}$ cm$^{-2}$, is in the 72-138 range, and in the 644-1308 range if the exciton density is assumed to be $10^{11}$ cm$^{-2}$. Therefore, the correlations in the positions of the electrons and the holes as a result of the attractive Coulomb interaction make electrons and holes in tightly bound excitons in TMDs far more susceptible to capture by defects compared to uncorrelated free carriers. Interestingly, even when the exciton density $n_{s,s',\alpha}$ is zero the capture rate in (\ref{eq:Rec_2}) is enhanced by the factors $G_{\alpha}$ compared to the rate in (\ref{eq:Rec_3}) for uncorrelated electrons and holes. Therefore, Coulomb correlations in the positions of electrons and holes due to the attractive interaction between them enhances the Auger scattering rates even at the Hartree-Fock level.

\section{Numerical Results and Discussion}

\subsection{Carrier Capture Times at Low Exciton Densities} \label{sec:num1}
\begin{figure}[tbh]
  \centering
  \includegraphics[width=.45\textwidth]{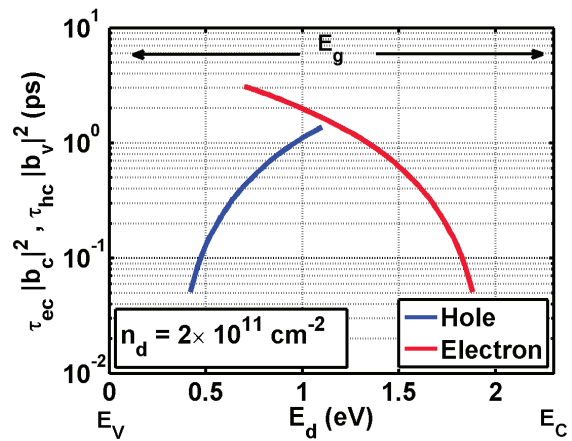}
  \caption{The capture times of electrons ($\tau_{ec}$) and holes ($\tau_{hc}$) of excitons by defects in monolayer MoS$_{2}$ on quartz are plotted as a function of the defect energy within the bandgap. The exciton binding energy is $E_{\alpha=1}$ is 0.4 eV and the material bandgap is 2.3 eV~\cite{Changjian14}. The plotted capture times for electrons and holes have been normalized by multiplying them by $|b_{c}|^{2}$ and $|b_{v}|^{2}$, respectively. The defect density $n_{d}$ is $2\times 10^{11}$ cm$^{-2}$.}
  \label{fig:capture1}
\end{figure}

For numerical computations, we consider monolayer MoS$_{2}$ on a quartz substrate, as is the case in many experiments. We first assume that the exciton density is small enough ($\le$$10^{12}$ cm$^{-2}$) to allow one to ignore phase-space filling effects~\cite{Changjian14}. We use the wavevector-dependent dielectric constant $\epsilon(\vec{q})$ for monolayer MoS$_{2}$ on quartz given by Zhang et~al.~\cite{Changjian14}. The defect state wavefunction is given in (\ref{eq:defect2}). The values of $|b_{c,s}|^{2}$ and $|b_{v,s}|^{2}$ are assumed to be independent of the valley/spin indices. This is a good approximation for many important cases. For example, in the case of the sulfur vacancy in MoS$_{2}$ discussed earlier, the $E$ states have a total weight of $\sim$0.25 on the $d_{z^{2}}$ orbitals of the Mo atoms adjacent to the missing sulfur atom~\cite{Yong14}. Since the conduction band Bloch states of both $K$ and $K'$ valleys are made up of mostly the $d_{z^{2}}$ orbitals of Mo atoms, $|b_{c,s}|^{2}$ is the same for both the valleys. We approximate the envelope, $\chi_{d}(\vec{r})$, of the defect state wavefunction in (\ref{eq:defect2}) by a Gaussian, $\chi_{d}(\vec{r}) = \sqrt{2/(\pi a_{d}^{2})}e^{-r^{2}/a_{d}^{2}}$, where $a_{d} \approx 3$ $\AA$ (see Fig.\ref{fig:s_vac}). Note that the in-plane S-Mo bound length in MoS$_{2}$ is $\sim$1.83 $\AA$. Fig.\ref{fig:capture1} plots the computed capture times of electrons ($\tau_{ec}$) and holes ($\tau_{hc}$) of excitons assuming that all the excitons are in the lowest state ($\alpha=1$). In the low exciton density limit considered here these capture times are independent of the exciton density. The defect density $n_{d}$ is assumed to be $2\times 10^{11}$ cm$^{-2}$. The capture times for electrons and holes shown in Fig.\ref{fig:capture1} have been normalized by multiplying them by $|b_{c}|^{2}$ and $|b_{v}|^{2}$, respectively, given the uncertainty in the exact values of these parameters. In the calculation of the electron capture times the defect state is assumed to be empty ($f_{d}=0$), and in the calculation of the hole capture times the defect state is assumed to be full ($f_{d}=1$).     

The curves shown in Fig.\ref{fig:capture1} can provide results in different situations. For example, in the case of the $E$ states associated with a sulfur vacancy, if $|b_{c}|^{2}$ is assumed to be $\sim$0.25~\cite{Yong14}, then the electron capture time curve in Fig.\ref{fig:capture1} would need to be multiplied by 4 in order to get the actual electron capture times. If the $E$ state energy is assumed to $\sim$1.5 eV above the valence band edge~\cite{Noh14}, then the electron capture time comes out to be $\sim$2.4 ps. Since the capture times decrease inversely with the defect density $n_{d}$, the capture times shown in Fig.\ref{fig:capture1} can be interpolated for different values of the defect density. For example, a defect density of $8\times 10^{11}$ cm$^{-2}$ would result in an electron capture time of 0.6 ps for the $E$ state of a sulfur vacancy (under the same assumptions as stated above).  

Fig.\ref{fig:capture1} shows that shallower traps have much shorter capture times than deeper traps. This can be understood as follows. Energy conservation requires that the scattered electron (hole), in a hole (electron) capture process, takes away most of the energy. The deeper the trap the more the final energy of the scattered particle. Also, momentum conservation requires that the momentum of the scattered particle be provided by the relevant Fourier component of the defect state wavefunction. Therefore, the deeper the trap the larger the momentum transfer. Since in Fourier space the defect state wavefunction is $\chi_{d}(\vec{q}) = \sqrt{2\pi a_{d}^{2}} e^{-q^{2}a_{d}^{2}/4}$, larger momentum transfers result in smaller capture rates. Note that this result is largely independent of the exact assumed form of the defect state wavefunction. In addition, the Coulomb potential $V(\vec{q})$ also decreases for larger momentum transfers. Although the final density of states available to the scattered particle increases with the particle energy (for non-parabolic energy band dispersions in 2D), this increase is not enough to offset the reduction in the capture rates due to the factors mentioned above.

Since the energy width of the valence and conduction bands in MoS$_{2}$ are less than 1.2 eV and 0.6 eV~\cite{Lam12,Louie1}, respectively, the limited horizontal extents of the curves in Fig.\ref{fig:capture1} ensure that the electron (hole) scattered to a high energy in the hole (electron) capture process is scattered within the same band consistent with the assumptions made in this work. It is, however, possible for the scattered particle to go into a different band. For example, slightly away from the $K$ ($K'$) points, the next higher conduction band has Bloch states with a large weight on the $d_{z^{2}}$ orbitals of Mo atoms and these Bloch states will have large overlap with the Bloch states near the conduction band bottom~\cite{Guinea13}. It should also be noted that the weights $|b_{c}|^{2}$ and $|b_{v}|^{2}$ for defects could be very small or zero. For example, in the case of sulfur vacancy $A_{1}$ states both $|b_{c}|^{2}$ and $|b_{v}|^{2}$ are expected to be very small~\cite{Robertson13,Noh14,Yong14}.

\begin{figure}[tbh]
  \centering
  \includegraphics[width=.45\textwidth]{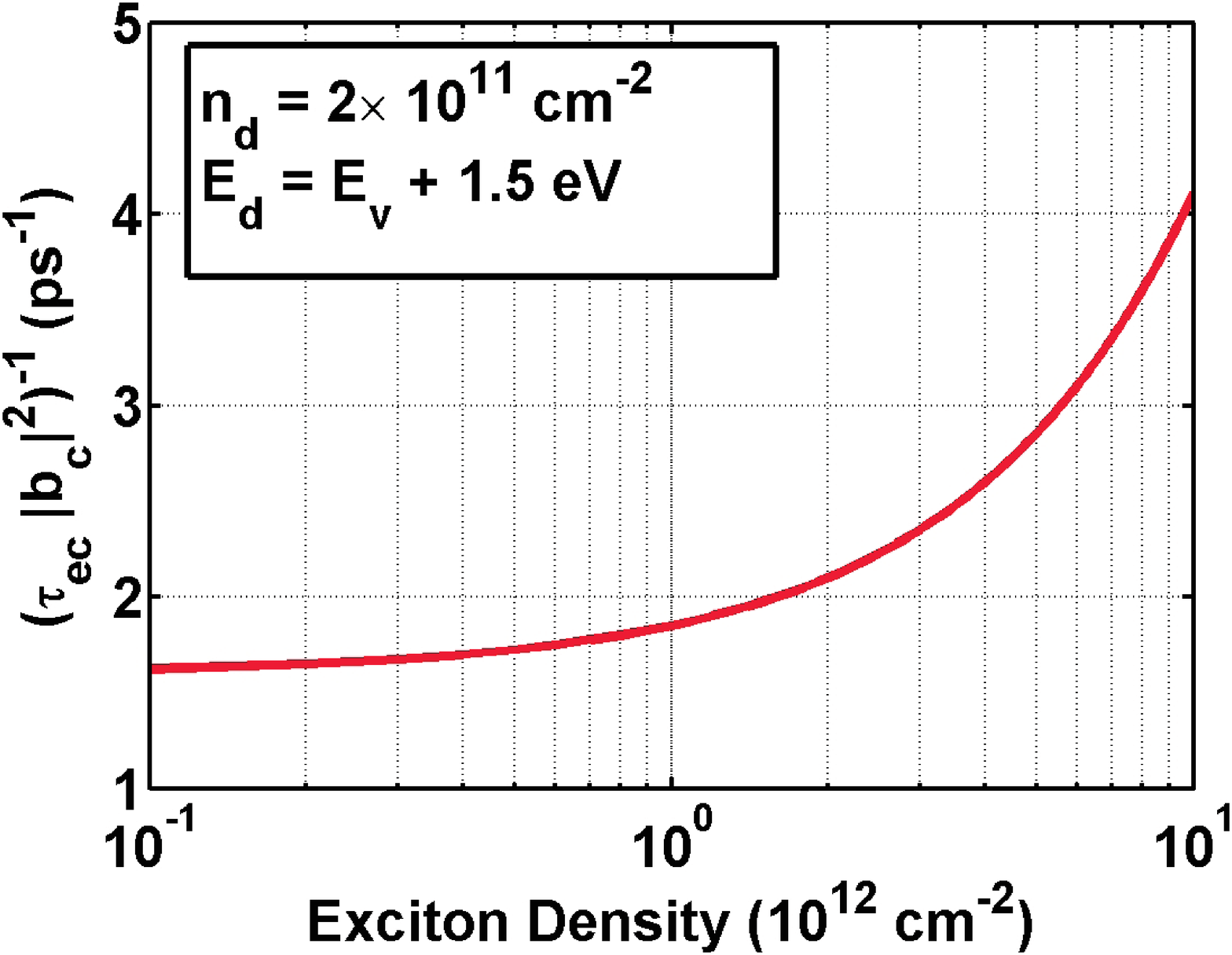}
  \caption{The inverse capture time ($\tau^{-1}_{ec}$) for the electron of an exciton in monolayer MoS$_{2}$ on quartz is plotted as a function of the exciton density. The plotted capture time has been normalized by multiplying it by $|b_{c}|^{2}$. The defect density $n_{d}$ is $2\times 10^{11}$ cm$^{-2}$ and the defect energy $E_{d}$ is assumed to be 1.5 eV above the valence band edge. The inverse capture time increases with the exciton density $n_{ex}$ roughly as, $\tau^{-1}_{ec} \sim A + Bn_{ex}$ ($A$ and $B$ are constants).}
  \label{fig:capture2}
\end{figure}

\subsection{Carrier Capture Times at High Exciton Densities}  \label{sec:num2}
At large exciton densities (typically larger than $10^{12}$ cm$^{-2}$, but smaller than $10^{13}$ cm$^{-2}$, for 2D-TMDs~\cite{Changjian14}), phase-space filling effects cannot be ignored in the description of the exciton states. We use the formalism developed by Kira and Koch~\cite{Kira12,Kira06}. When phase-space filling is taken into account, exciton eigenvalue equation in the relative co-ordinates becomes non-Hermitian (see the Appendix) and its solutions are expressed in terms of the {\em left} and the {\em right} eigenfunctions, $\phi^{L}_{\alpha,s,s'}(\vec{k},\vec{Q})$ and $\phi^{R}_{\alpha,s,s'}(\vec{k},\vec{Q})$, respectively. These eigenfunctions are a also a function of the center of mass momentum $\vec{Q}$, and are related as follows~\cite{Kira12,Kira06},
\begin{eqnarray}
& & \phi^{R}_{s,s',\alpha}(\vec{k},\vec{Q}) = \phi^{L}_{s,s',\alpha}(\vec{k},\vec{Q}) \left[ f_{v,s'}(\vec{k} - \frac{m_{h}}{m_{ex}} \vec{Q}) \right. \nonumber \\ 
& & \left. - f_{c,s}(\vec{k} + \frac{m_{e}}{m_{ex}} \vec{Q}) \right] 
\end{eqnarray}
and obey the orthogonality relation,
\begin{equation}
\int \frac{d^{2}\vec{k}}{(2\pi)^{2}} [\phi^{L}_{s,s',\alpha}(\vec{k},\vec{Q})]^{*} \phi^{R}_{s,s',\beta}(\vec{k},\vec{Q}) = \delta_{\alpha,\beta} \label{eq:orth_2} 
\end{equation}
In terms of these eigenfunctions, the expression for the electron capture rate becomes,
\begin{eqnarray}
& & R_{ec} \approx \frac{2\pi}{\hbar} n_{d}(1-f_{d}) \frac{1}{A} \sum_{s,s',\alpha,\vec{Q}} D_{v,s'}(q_{\alpha}) |\chi_{d}(q_{\alpha})|^{2}  |b_{c,s}|^{2}   \nonumber \\ 
& & \times \left| \frac{1}{A} \sum_{\vec{k}_{r}} V(q_{\alpha}\hat{x}-\vec{k}_{r})  \phi^{R}_{s,s',\alpha}(\vec{k}_{r},\vec{Q}) \right|^{2} \left[ n_{s,s',\alpha}(\vec{Q}) \right. \nonumber \\
& & \left. + \frac{1}{A} \sum_{\vec{k}} |\phi^{L}_{s,s',\alpha}(\vec{k},\vec{Q})|^{2} f_{c,s}(\vec{k} + \frac{m_{e}}{m_{ex}} \vec{Q}) \right. \nonumber \\
& & \left. \times \left[ 1 - f_{v,s'}(\vec{k} -  \frac{m_{h}}{m_{ex}} \vec{Q}) \right] \right] \nonumber \label{eq:Rec_4} \\ 
\end{eqnarray}  
The expression for the capture rate of holes in the high exciton density case follows similarly from (\ref{eq:Rhc_0}). When all electrons and holes exist as excitons, self-consistency requires that the distribution functions are given by~\cite{Kira12},
\begin{eqnarray}
& & f_{c,s}(\vec{k}) = \frac{1}{A} \sum_{s',\alpha,\vec{Q}} [\phi^{L}_{s,s',\alpha}(\vec{k},\vec{Q})]^{*} \phi^{R}_{s,s',\alpha}(\vec{k},\vec{Q}) n_{s,s',\alpha}(\vec{Q}) \nonumber \\
& & 1 - f_{v,s'}(\vec{k}) = \frac{1}{A} \sum_{s,\alpha,\vec{Q}} [\phi^{L}_{s,s',\alpha}(\vec{k},\vec{Q})]^{*} \phi^{R}_{s,s',\alpha}(\vec{k},\vec{Q}) n_{s,s',\alpha}(\vec{Q}) \nonumber \label{eq:dist} \\
\end{eqnarray}
Equations (\ref{eq:Rec_4}) and (\ref{eq:dist}) show that the capture rate $R_{ec}$ has terms that go linearly as well as quadratically with the exciton density. The quadratic dependence comes from the Hartree-Fock term in the evaluation of $\langle B^{\dagger}_{s,s',\alpha}(\vec{Q})  B_{s,s',\alpha}(\vec{Q}) \rangle$ (see Equation (\ref{eq:ex_avg})). It can be understood as coming from the Auger scattering between the electron of one exciton and the hole of another exciton. Recall from the discussion in Section \ref{sec:corr} that even at the Hartree-Fock level Auger scattering between electrons and holes is enhanced due to the Coulomb correlations compared to uncorrelated electrons and holes.

For numerical computations, we again consider monolayer MoS$_{2}$ on a quartz substrate, as in Section \ref{sec:num1}. We solve the exciton eigenvalue equation for different exciton densities and obtain the exciton radii and the exciton binding energies~\cite{Changjian14}. For simplicity, we consider the case when all the electrons and holes are in the lowest ($\alpha =1$) bound exciton state. Fig.\ref{fig:capture2} plots the inverse capture time ($\tau^{-1}_{ec}$) of the electron of an exciton in monolayer MoS$_{2}$ on quartz as a function of the exciton density. The plotted capture time has been normalized by multiplying it by $|b_{c}|^{2}$. The defect density $n_{d}$ is $2\times 10^{11}$ cm$^{-2}$ and the defect energy $E_{d}$ is assumed to be 1.5 eV above the valence band edge. The inverse capture time increases with the exciton density $n_{ex}$ roughly as, $\tau^{-1}_{ec} \sim A + Bn_{ex}$ ($A$ and $B$ are constants), indicating that the capture rate $R_{ec}$ has both linear and quadratic dependence on the exciton density ($R_{ec} \sim A n_{ex} + B n_{ex}^{2}$). The term quadratic in the exciton density in $R_{ec}$ becomes significant at exciton densities higher than $\sim$10$^{12}$ cm$^{-2}$. When interpreting experimental data, this quadratic increase of the carrier capture rate with the exciton density can make exciton annihilation via carrier capture by defects indistinguishable from direct electron-hole recombination via interband Auger scattering (exciton-exciton annihilation), the rate of which is also expected to go quadratically with the exciton density.

\section{Comments and Conclusion}
The results presented in this paper show that the capture times for electrons and holes of excitons in TMDs can be very short - from less than a picosecond to a few picoseconds. These numbers agree well with the recently reported experimental results on the ultrafast carrier dynamics in photoexcited monolayer MoS$_{2}$ where fast relaxation times in the few picoseconds range were observed~\cite{Shi13,Korn11,Lagarde14,Wangb14}. In addition, the results in Fig.\ref{fig:capture1} and Fig.\ref{fig:capture2} are largely independent of the carrier temperature which is also consistent with the experimental observations~\cite{Lagarde14,Wangb14}. 

The expressions given in this work could overestimate (underestimate) the capture rates (times). The reasons are as follows. The magnitude of the intraband overlap integrals for Bloch states were assumed to equal unity in Section \ref{sec:Hc} and only phase differences were taken into account. At energies much different from the band edge energies, the Bloch states are different from the band edge Bloch states, and consequently the magnitude of the overlap integrals are smaller than unity. For example, the two-band $k.p$ model in Section \ref{sec:nonintH} shows that at wavevector $\vec{k}$ the conduction (valence) band Bloch states have contributions from the valence (conduction) band Bloch states at $\vec{k}=0$ with a weight given by $0.5 - 0.5 (\Delta_{s}/2)/\sqrt{(\Delta_{s}/2)^2 + (\hbar v k)^{2}}$. This implies a 15\% weight at energies in the band that are $\sim$0.5 eV away from the band edge. In addition, both the conduction and valence band Bloch states are expected to get contributions from other lower and higher bands at large wavevectors~\cite{Falko13}. However, we don't expect the essential physics to change significantly or the rates to change by more than a factor of unity when these sources of error are removed. We should also point out that the rates for carrier capture by defects in 2D-TMDs can vary from sample to sample as the nature of defects is expected to depend on the method of sample preparation.

\section{Acknowledgments}
The authors would like to acknowledge helpful discussions with Paul L. McEuen and Michael G. Spencer, and support from CCMR under NSF grant number DMR-1120296, AFOSR-MURI under grant number FA9550-09-1-0705, and ONR under grant number N00014-12-1-0072.

\section{Appendices} 

\subsection{Details on the Electron Capture Rate}
In this Section, we derive the expression for the electron capture rate given in (\ref{eq:Rec_0}). The derivation of the hole capture rate is similar. We assume an initial state described by the density matrix $\rho_{i}$ in which the exciton occupation is $n_{s,s',\alpha}(\vec{Q})$, the defect density is $n_{d}$, the defect occupation is $f_{d}$, and the electron and hole densities (including both free carriers and bound excitons) are $n_{s}$ and $p_{s'}$, respectively. The average values of various operators are as given in (\ref{eq:ex_avg}). The rate of change of the total electron density is,
\begin{equation}
\dot{n} = \frac{dn}{dt} = \frac{d}{dt} \left( \frac{1}{A}\sum_{s,\vec{k}} c^{\dagger}_{\vec{k},s} c_{\vec{k},s} \right)
\end{equation}
Defining the interaction representation for the time development of operators as,
\begin{equation}
O^{I}(t) = e^{\frac{i}{\hbar}(H_{o}+H_{eh}) t} O e^{-\frac{i}{\hbar}(H_{o}+H_{eh}) t}
\end{equation}
the rate $R_{ec}$ for the electron capture by the defect can be found by picking the appropriate term from the expression obtained using the first order perturbation theory,
\begin{eqnarray}
\langle \frac{dn}{dt} \rangle = \lim_{\eta \rightarrow 0} \frac{i}{\hbar}An_{d}\int_{-\infty}^{t} dt' \, e^{\eta t'} \, {\rm Tr}\left\{ \rho_{i} \left[ H_{ec}^{I}(t') , \dot{n}^{I}(t) \right] \right\} \nonumber \\
\end{eqnarray}
Since the exciton states are approximate eigenstates of the Hamiltonian $H_{o} + H_{eh}$ we have,
\begin{equation}
e^{\frac{i}{\hbar}(H_{o}+H_{eh}) t} B_{s,s',\alpha}(\vec{Q}) e^{-\frac{i}{\hbar}(H_{o}+H_{eh}) t} \approx  B_{s,s',\alpha}(\vec{Q}) e^{-i\frac{E_{\alpha}(\vec{Q})}{\hbar}t}
\end{equation}
It is therefore convenient to express the conduction and valence band creation and destruction operators appearing in $H_{ec}$ using the exciton basis described in Section~\ref{sec:ex_basis}. We also point out here that the ensemble average of a product of operators of the form,
\begin{equation}
\langle e^{\frac{i}{\hbar}(H_{o}+H_{eh}) t} c^{\dagger}_{\vec{k}_{1},s_{1}} b_{\vec{k}'_{1},s'_{1}} b^{\dagger}_{\vec{k}'_{2},s'_{2}} c_{\vec{k}_{2},s_{2}} e^{-\frac{i}{\hbar}(H_{o}+H_{eh}) t} \rangle 
\end{equation}
needs to be evaluated using the cluster expansion and keeping the correlation terms as well as the Hartree-Fock term~\cite{Kira06,Koch06}. The final result is,
\begin{eqnarray}
& & R_{ec} = \frac{2\pi}{\hbar} n_{d}(1-f_{d}) \frac{1}{A^{4}} \mathop{\sum_{s,s',\vec{k}_{r},\vec{k}'_{r}}}_{\vec{Q},\vec{q},\alpha} |b_{c,s}|^{2} V^{*}(\vec{q}-\vec{k}'_{r})  V(\vec{q}-\vec{k}_{r}) \nonumber \\
& & \times |\chi_{d}(\vec{q} + (m_{e}/m_{ex})\vec{Q})|^{2} \phi^{*}_{\alpha}(\vec{k}'_{r}) \phi_{\alpha}(\vec{k}_{r}) \left[ n_{s,s',\alpha}(\vec{Q}) \right. \nonumber \\
& & \left. + \frac{1}{A} \sum_{\vec{k}} |\phi_{\alpha}(\vec{k})|^{2} f_{c,s}(\vec{k} + \frac{m_{e}}{m_{ex}} \vec{Q}) \left[ 1 - f_{v,s'}(\vec{k} -  \frac{m_{h}}{m_{ex}} \vec{Q}) \right] \right] \nonumber \\
& & \times \delta \left( E_{g} - E_{\alpha} + \frac{\hbar^{2}Q^{2}}{2m_{ex}} - E_{d} \right. \nonumber \\
& & \left. - E_{v,s'}(0) + E_{v,s'}(\vec{q} - \frac{m_{h}}{m_{ex}}\vec{Q}) \right)
\end{eqnarray}
Note that all the phase factors have canceled out. The exciton center of mass kinetic energy, $\hbar^{2}Q^{2}/2m_{ex}$, is expected to be much smaller than the energy difference $E_{g} - E_{\alpha} - E_{d}$. The former is expected to be in the few tens of meV range and the latter in the hundreds of meV range. The energy conserving delta function then enforces $q$ to the value determined by the condition $E_{v,s'}(0) - E_{v,s'}(q_{\alpha}) = E_{g} - E_{\alpha} - E_{d}$. Once the magnitude of $\vec{q}$ has been fixed in this way, it is easy to see that $R_{ec}$ does not depend on the angle of $\vec{q}$. So one may assume $q \approx q_{\alpha} \hat{x}$ and obtain,
\begin{eqnarray}
& & R_{ec} \approx \frac{2\pi}{\hbar} n_{d}(1-f_{d}) \sum_{s,s',\alpha} D_{v,s'}(q_{\alpha}) |\chi_{d}(q_{\alpha})|^{2}  |b_{c,s}|^{2}   \nonumber \\ 
& & \times \left| \frac{1}{A} \sum_{\vec{k}_{r}} V(q_{\alpha}\hat{x}-\vec{k}_{r})  \phi_{\alpha}(\vec{k}_{r}) \right|^{2} \left[ n_{s,s',\alpha} \right. \nonumber \\
& & \left. + \frac{1}{A^{2}} \sum_{\vec{k},\vec{Q}} |\phi_{\alpha}(\vec{k})|^{2} f_{c,s}(\vec{k} + \frac{m_{e}}{m_{ex}} \vec{Q}) \left[ 1 - f_{v,s'}(\vec{k} -  \frac{m_{h}}{m_{ex}} \vec{Q}) \right] \right] \nonumber \\ 
\end{eqnarray}
Here, $D_{v,s'}$ is the valence band density of states (per valley per spin) evaluated at the energy of the scattered hole whose wavevector is $q_{\alpha}$. The exciton density $n_{s,s',\alpha}$ is, 
\begin{equation}
n_{s,s',\alpha} = \int \frac{d^{2}\vec{Q}}{(2\pi)^{2}} \, n_{s,s',\alpha}(\vec{Q})
\end{equation} 
If $q_{\alpha} >> k_{r}$ for all values of $k_{r}$ for which $\phi_{\alpha}(\vec{k}_{r})$ is significant, then the above expression reduces to, 
\begin{eqnarray}
R_{ec} & = & \frac{2\pi}{\hbar} n_{d} \left( 1-f_{d} \right) \sum_{s,s',\alpha} D_{v,s'}(q_{\alpha}) |V(q_{\alpha})|^{2} |\chi_{d}(q_{\alpha})|^{2}  \nonumber \\
& &  \times |b_{c,s}|^{2} \left[ |\phi_{\alpha}(\vec{r}=0)|^{2} n_{s,s',\alpha} + G_{\alpha} n_{s}p_{s'} \right]         \label{eq:Rec_app}
\end{eqnarray}
Equation (\ref{eq:Rec_app}) contains the exciton density $n_{s,s',\alpha}$ as well as the electron and hole densities (including both free carriers and bound excitons) $n_{s}$ and $p_{s'}$, respectively. The latter appear as a result of the Hartree-Fock term in the cluster expansion~\cite{Kira06,Koch06}. $G_{\alpha}$ is,
\begin{eqnarray}
&& G_{\alpha} = \frac{|\phi_{\alpha}(\vec{r}=0)|^{2}}{n_{s}p_{s'} \, A^{2}} \sum_{\vec{k},\vec{Q}} |\phi_{\alpha}(\vec{k})|^{2} f_{c,s}(\vec{k} + \frac{m_{e}}{m_{ex}} \vec{Q})  \nonumber \\
&& \times  \left[ 1 - f_{v,s'}(\vec{k} -  \frac{m_{h}}{m_{ex}} \vec{Q}) \right]
\end{eqnarray}
$G_{\alpha}$ is expected to be significant for only the lowest few exciton states. 

If all the electrons and holes are assumed to be in the lowest ($\alpha =1$) bound exciton state then self-consistency requires that the distribution functions are given by~\cite{Kira12},
\begin{eqnarray}
f_{c,s}(\vec{k}) & = &  |\phi_{\alpha=1}(\vec{k})|^{2} n_{s} \nonumber \\
1 - f_{v,s'}(\vec{k}) & = &  |\phi_{\alpha=1}(\vec{k})|^{2}  p_{s'} \label{eq:dist_2}
\end{eqnarray}
Here, $n_{s} = \sum_{s'} n_{s,s',\alpha=1}$ and $p_{s'} = \sum_{s} n_{s,s',\alpha=1}$. One then obtains,
\begin{eqnarray}
&& G_{1} = |\phi_{\alpha=1}(\vec{r}=0)|^{2} \frac{1}{A^{2}} \sum_{\vec{k},\vec{Q}}  |\phi_{\alpha=1}(\vec{k} + \frac{m_{e}}{m_{ex}} \vec{Q})|^{2} \nonumber \\
&&  \times |\phi_{\alpha=1}(\vec{k})|^{2} |\phi_{\alpha=1}(\vec{k} - \frac{m_{h}}{m_{ex}} \vec{Q})|^{2}
\end{eqnarray}
Assuming the standard 2D exciton wavefunction~\cite{Changjian14}, $G_{1}$ equals $128/(5\pi) \approx 8.15$.

\subsection{Description of Excitons States in the High Exciton Density Limit}
In the high exciton density case, phase filling effects cannot be ignored in the description of the exciton states~\cite{Changjian14,Kira12}. The exciton wavefunctions, $\phi^{L}_{\alpha,s,s'}(\vec{k},\vec{Q})$ and $\phi^{R}_{\alpha,s,s'}(\vec{k},\vec{Q})$ satisfy the eigenvalue equations~\cite{Kira12},
\begin{eqnarray}
&& \left[ E_{c,s}(\vec{k}+(m_{e}/m_{ex})\vec{Q}) - E_{v,s'}(\vec{k}-(m_{h}/m_{ex})\vec{Q}) \right] \nonumber \\
&& \times \phi^{L}_{\alpha,s,s'}(\vec{k},\vec{Q}) - \frac{1}{A}\sum_{\vec{k}'} V(\vec{k}-\vec{k}') \phi^{L}_{\alpha,s,s'}(\vec{k}',\vec{Q}) \nonumber \\
&& \times \left[ f_{v,s'}(\vec{k}' - \frac{m_{h}}{m_{ex}} \vec{Q}) - f_{c,s}(\vec{k}' + \frac{m_{e}}{m_{ex}} \vec{Q}) \right]  \nonumber \\
&& = E_{s,s',\alpha}(\vec{Q})  \phi^{L}_{s,s',\alpha}(\vec{k},\vec{Q})
\end{eqnarray}
\begin{eqnarray}
&& \left[ E_{c,s}(\vec{k}+(m_{e}/m_{ex})\vec{Q}) - E_{v,s'}(\vec{k}-(m_{h}/m_{ex})\vec{Q}) \right] \nonumber \\
&& \times \phi^{R}_{\alpha,s,s'}(\vec{k},\vec{Q}) - \left[ f_{v,s'}(\vec{k} - \frac{m_{h}}{m_{ex}} \vec{Q}) \right. \nonumber \\
&& \left. - f_{c,s}(\vec{k} + \frac{m_{e}}{m_{ex}} \vec{Q}) \right] \frac{1}{A}\sum_{\vec{k}'} V(\vec{k}-\vec{k}') \phi^{R}_{\alpha,s,s'}(\vec{k}',\vec{Q}) \nonumber \\
&& = E_{s,s',\alpha}(\vec{Q})  \phi^{R}_{s,s',\alpha}(\vec{k},\vec{Q})
\end{eqnarray}
The exciton wavefunctions satisfy the orthogonality and completeness relations,
\begin{equation}
\int \frac{d^{2}\vec{k}}{(2\pi)^{2}} [\phi^{L}_{s,s',\alpha}(\vec{k},\vec{Q})]^{*} \phi^{R}_{s,s',\beta}(\vec{k},\vec{Q}) = \delta_{\alpha,\beta} \label{eq:orth_app} 
\end{equation}
\begin{equation}
\sum_{\alpha} \phi^{L}_{s,s',\alpha}(\vec{k},\vec{Q}) [\phi^{R}_{s,s',\alpha}(\vec{k}',\vec{Q})]^{*} = (2\pi)^{2} \delta^{2}(\vec{k}-\vec{k}') \label{eq:comp_app}
\end{equation}
We also define,
\begin{eqnarray}
 \psi^{L/R}_{s,s',\alpha}(\vec{k},\vec{Q}) & = & \phi^{L/R}_{s,s',\alpha}(\vec{k},\vec{Q})  \nonumber \\
&  \times & e^{i(\tau\phi_{\vec{k}+ (m_{e}/m_{ex})\vec{Q}} + \tau'\phi_{\vec{k}'-(m_{h}/m_{ex})\vec{Q}})/2}
\end{eqnarray}
The exciton creation operator $B^{\dagger}_{s,s',\alpha}(\vec{Q})$ is defined as,
\begin{equation}
 B^{\dagger}_{s,s',\alpha}(\vec{Q}) =  \frac{1}{\sqrt{A}}\sum_{\vec{k}} \psi^{L}_{s,s',\alpha}(\vec{k},\vec{Q}) c_{\vec{k} + \frac{m_{e}}{m_{ex}}\vec{Q},s}^{\dagger} b_{\vec{k}-\frac{m_{h}}{m_{ex}}\vec{Q},s'} 
\end{equation}
Using the completeness and the orthogonality of the exciton wavefunctions given in (\ref{eq:comp_app}) and (\ref{eq:orth_app}), we get,
\begin{equation}
c_{\vec{k},s}^{\dagger} b_{\vec{k}',s'} = \frac{1}{\sqrt{A}} \sum_{\alpha} \psi^{R}_{s,s',\alpha}(\vec{k}_{r},\vec{Q}) B^{\dagger}_{s,s',\alpha}(\vec{Q})
\end{equation}
where, $\vec{k}_{r}$ and $\vec{Q}$ equal $(m_{h}/m_{ex})\vec{k} + (m_{e}/m_{ex})\vec{k}'$ and $\vec{k}-\vec{k}'$ on the left hand side, respectively.


\begin{thebibliography}{99}

\bibitem{Fai10} K. F. Mak, C. Lee, J. Hone, J. Shan, and T. F. Heinz, Phys. Rev. Lett. 105, 136805 (2010).
\bibitem{Xu13}  J. S. Ross, S. Wu, H. Yu, N. J. Ghimire, A. M. Jones, G. Aivazian, J. Yan, D. G. Mandrus, D. Xiao, W. Yao, and X. Xu, Nat. Comm. 4, 1474 (2013). 	
\bibitem{Changjian14} C. Zhang, H. Wang, W. Chan, C. Manolatou, F. Rana, Phys. Rev. B, 89, 205436 (2014). 
\bibitem{timothy} T. C. Berkelbach, M. S. Hybertsen, and D. R. Reichman, Phys. Rev. B 88, 045318 (2013).
\bibitem{Chernikov14} A. Chernikov, T. C. Berkelbach, H. M. Hill, A. Rigosi, Y. Li, O. B. Aslan, D. R. Reichman, M. S. Hybertsen, T. F. Heinz, Phys. rev. Lett., 113, 076802 (2014).
\bibitem{Efimkin13} D. K. Efimkin, Yu. E. Lozovik, Phys. Rev. B, 87, 245416 (2013).
\bibitem{Konabe14} S. Konabe, S. Okada, Phys. Rev. B, 90, 15304 (2014).
\bibitem{Berg14} G. Berghauser, E. Malic, Phys. Rev. B, 89, 125309 (2014).  


\bibitem{Liu14} G. Liu, W. Shan, Y. Yao, W. Yao, D. Xiao, Phys. Rev., B, 88, 085433 (2014).   
\bibitem{Wanga14} H. Wang, C. Zhang, W. Chan, C. Manolatou, S. Tiwari, F. Rana, arXiv:1409.3996 (2014). 
\bibitem{Landsberg92} P. T. Landsberg, ``Recombination in Semiconductors'', Cambridge University Press, Cambridge, UK (1992). 
\bibitem{Landsberg80} D J Robbins, P T Landsberg, Journal of Physics C, 13, 2425 (1980).
\bibitem{Shi13} H. Shi, R. Yan, Rusen, S. Bertolazzi, J. Brivio, B. Gao, A. Kis, D. Jena, H. Xing, Huili, L. Huang, ACS Nano, 7, 1072 (2013).
\bibitem{Lagarde14} D. Lagarde, L. Bouet, X. Marie, C. R. Zhu, B. L. Liu, T. Amand, P H. Tan, B. Urbaszek, Phys. Rev. Lett., 112, 047401 (2014).
\bibitem{Korn11} T. Korn, S. Heydrich, M. Hirmer, J. Schmutzler, C. Schüller, App. Phys. Lett., 99, 102109 (2011).  
\bibitem{yao12} D. Xiao, Gui-Bin Liu, W. Feng, X. Xu, and W. Yao, Phys. Rev. Lett. 108, 196802 (2012).


\bibitem{Yong14} Private communication with Yong-Sung Kim~\cite{Noh14}. 
\bibitem{Guinea13} E. Cappelluti, R. Roldan, J. A. Silva-Guillen, P. Ordejon, F. Guinea, Phys. Rev., B, 88, 075409 (2013).
\bibitem{Kira06} M. Kira, S. W. Koch, Prog. Quant. Electron., 30, 155 (2006). 
\bibitem{Koch06} S. W. Koch, M. Kira, G. Khitrova, H. M. Gibbs, Nature Materials, 5, 523 (2006).
\bibitem{Lam12} T. Cheiwchanchamnangij and W. R. L. Lambrecht, Phys. Rev. B 85, 205302 (2012).
\bibitem{Louie1} D. Y. Qiu, F. H. da Jornada, and S. G. Louie, Phys. Rev. Lett. 111, 216805 (2013).
\bibitem{Wangb14} H. Wang, C. Zhang, F. Rana,arXiv:1409.4518 (2014). 
\bibitem{Kira12} M. Kira, S. W. Koch, ``Semiconductor Quantum Optics'', Cambridge University Press, NY (2012).  
\bibitem{Lopez13} O. Lopez-Sanchez, D. Lembke, M. Kayci, A. Radenovic, and A. Kis, Nature Nanotechnology 8, 497 (2013).
\bibitem{Ross14} J. S. Ross, P. Klement, A. M. Jones, N. J. Ghimire, J. Yan, D. G. Mandrus, T. Taniguchi, K. Watanabe, K. Kitamura, W. Yao, et~al., Nature Nanotechnology 9, 268 (2014).
\bibitem{Yin12} Z. Yin, H. Li, H. Li, L. Jiang, Y. Shi, Y. Sun, G. Lu, Q. Zhang, X. Chen, and H. Zhang, ACS Nano 6, 74 (2012).
\bibitem{Steiner13} R. S. Sundaram, M. Engel, A. Lombardo, R. Krupke, A. C. Ferrari, P. Avouris, and M. Steiner, Nano Letters 13, 1416 (2013). 
\bibitem{Pablo14} B. W. H. Baugher, H. O. H. Churchill, Y. Yang, and P. Jarillo-Herrero, Nature Nanotechnology 9, 262 (2014).
\bibitem{Feng10} A. Splendiani, L. Sun, Y. Zhang, T. Li, J. Kim, C.-Y. Chim, G. Galli, and F. Wang, Nano Letters, 10, 1271 (2010).
\bibitem{Falko13} A. Kormanyos, V. Zolyomi, N. D. Drummond, P. Rakyta, G. Burkard and V. I. Falko, Phys. Rev. B 88, 045416 (2013).
\bibitem{VanDerZande13} A. M. van der Zande, P. Y. Huang, D. A. Chenet, T. C. Berkelbach, Y. You, G.-H. Lee, T. F. Heinz, D. R. Reichman, D. A. Muller, and J. C. Hone, Nature Materials, 12, 554 (2013).

\bibitem{Sofo04} J. D. Fuhr, A. Saul, and J. O. Sofo, Phys. Rev. Lett. 92, 026802 (2004).
\bibitem{Komsa12} H. P. Komsa, J. Kotakoski, S. Kurasch, O. Lehtinen, U. Kaiser, and A. V. Krasheninnikov, Phys. Rev. Lett., 109, 035503 (2012).
\bibitem{Seifert13} A. N. Enyashin, M. Bar-Sadan, L. Houben, and G. Seifert, The Journal of Physical Chemistry C, 117, 10842 (2013).
\bibitem{Kong13} W. Zhou, X. Zou, S. Najmaei, Z. Liu, Y. Shi, J. Kong, J. Lou, P. M. Ajayan, B. I. Yakobson, J. C. Idrobo, Nano Lett., 13, 2615 (2013).
\bibitem{Noh14} J. Noh, H. Kim, Y. Kim, Phys. Rev., B, 89, 205417 (2014).
\bibitem{Robertson13} D. Liu, Y. Guo, L. Fang, J. Robertson, Appl. Phys. Lett., 103, 183113 (2013).   
\bibitem{Guinea14} S. Yuan, R. Roldan, M. I. Katsnelson, and F. Guinea, Phys. Rev. B 90, 041402 (2014).
\bibitem{Hao13} H. Qiu, T. Xu, Z. Wang, W. Ren, H. Nan, Z. Ni, Q. Chen, S. Yuan, F. Miao, F. Song, et al., Nature Communications, 4, 2642 (2013).

\end{thebibliography}
\end{document}